\documentclass[aps,preprint,onecolumn]{revtex4}
\usepackage{graphicx}
\usepackage{amssymb}
\usepackage{amsmath}
\usepackage{amsfonts}
\usepackage{graphicx}%
\begin{document}
\bibliographystyle{prsty}
\title{Extension of Bethe's diffraction model to conical Geometry :\\
application to near field optics}
\author{A.~Drezet, J.~C.~Woehl, and S.~Huant}
\affiliation{Laboratoire de Spectrométrie Physique, CNRS UMR5588, Universit\'e Joseph Fourier Grenoble,
BP 87, 38402 Saint Martin d'Hères cedex, France }

\date{\today}

\begin{abstract} The generality of the Bethe's two dipole model for light diffraction through a subwavelength aperture in a conducting plane is studied in the radiation zone for coated conical fiber tips as those used in near field scanning optical microscopy. In order to describe the angular radiated power of the tip
theoretically, we present a simple, analytical model for small apertures (radius $\lesssim$ 40 nm) based on a multipole
expansion. Our model is able to reproduce the available experimental results. It proves
relatively insensitive
 to cone angle and aperture radius and contains, as a first approximation, the empirical two-dipole model proposed earlier.
\end{abstract}

\maketitle

\begin{figure}
\includegraphics[width=10cm]{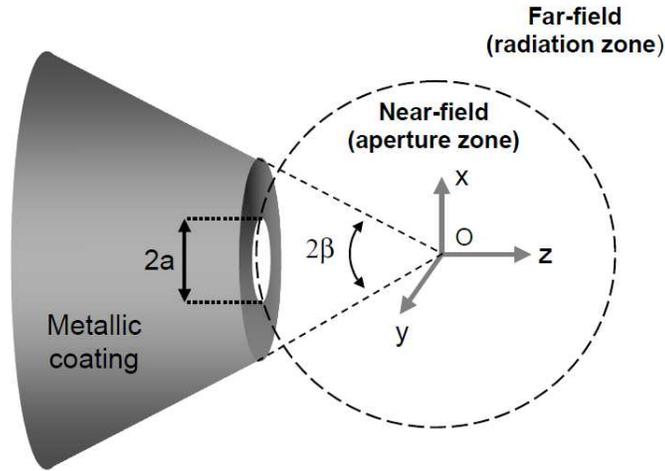} \caption{Geometry of a NSOM fiber tip. It is characterized by \textit{i}) the aperture radius $a$ in the metallic coating and \textit{ii}) the tip angle $\beta$. The quasi-multipole model is valid only in the far-field zone where $a/\left(\sin{\beta}\right)\leq r$.
 The boundary conditions cannot be applied in the near-field (or aperture) zone because it contains no metal.} \label{fig2}
\end{figure}

\emph{Introduction and experimental motivation}~-~
The well-established Fresnel and perturbation methods used in optics are not adapted to diffraction phenomena implying strong near-field patterns which occur for wavelengths $\lambda$ comparable to the dimensions of the diffracting object. This is due to the fact that for small objects, the ``interaction zone", of typical dimension $\lambda$, invades the space around  the diffracting matter. As a consequence, the interaction of light with matter cannot be considered as a small perturbation. Rayleigh\cite{1}, Bethe\cite{2}, and Bouwkamp\cite{3,4} have studied the particular case of diffraction by a small aperture in a perfect metallic screen. This problem~-~hereafter referred to as the Bethe's problem~-~is very typical because it contains dipolar terms which dominate in the ``radiation" far field of the hole. The present work addresses the validity and the survival of the diffraction phenomenology revealed by the Bethe's model in the case of small aperture at the apex of a conical screen.\\Such an unusual geometry is encountered in Near Field Scanning Optical Microscopy (NSOM). Indeed in NSOM, a conical optical tip with a
sub-wavelength aperture at the apex is raster-scanned in close proximity to the
investigated surface. Most NSOM make use of the tapered and coated optical fiber introduced some 10 years ago by Betzig \emph{et} \textsl{al.} \cite{5}. The
 angular radiation (far field) profile of the tip depends on the aperture diameter as has been extensively discussed by C.~Obermuller and
K.~Karrai \cite{6,7}. For small aperture sizes,
 the far-field
 radiation pattern is polarization-dependent with a large backward emission in the P-polarization in contrast to the S-polarization
(incident electric field parallel and perpendicular to the analysis plane, respectively) for which there is little backward emission. For small radiation angles, a simple combination of mutually perpendicular
electric and magnetic radiation dipoles situated in the aperture plane reproduces the main features of the angular radiation
 pattern for the smallest apertures (60 to 80 nm diameter) \cite{6,7}.
There is presently no formal justification for this 2-dipole model, and for that reason, the transmission profile of tapered optical probes, i.~e. the problem of light diffraction through a subwavelenght hole in a conical screen, needs further
theoretical examination. \\
%
%
In this letter, we solve Maxwell's equations analytically in the radiation zone with the use of appropriate boundary conditions in order to get insight into the peculiarities of a conical diffraction geometry compared to the more common planar geometry. As a critical test of our analytical model, we compare calculated emission profiles with available experimental profiles for both polarization states. Our analysis, based on a conical tip geometry with an aperture radius that is negligible compared to the wavelength of the incident light, is able to reproduce
the radiation pattern of small apertures over a large angular range. Remaining discrepancies at the largest radiation
angles can be accounted for by the boundary conditions in the metallic coating.
The above-mentioned empirical radiation dipole approach is formally justified for small radiation angles.
In addition, it is shown that the angular power spectrum does not critically depend on the exact tip geometry. Especially the
dependence on the taper angle is found to be very weak.\\
%
%
\emph{Scatterring and general properties}~-~In order to illustrate the main physical effects of small apertures in metallic surfaces, it is worthwhile to look at similar problems that have been treated historically.
As already mentionned in the introduction, Rayleigh\cite{1} and Bethe\cite{2} (see also Ref.~\cite{3}) have studied the light diffraction by a subwavelength
aperture in infinite, perfectly conducting screen. They have demonstrated that, in the far field, the
diffracted electromagnetic field can be described by a combination
of effective dipoles which depend on the local incident field
(${\mathbf E}_{\text{inc}}$,${\mathbf B}_{\text{inc}}$) and on the hole size
 $a$:
 \begin{eqnarray}
{\mathbf P}_{\text{eff}}= \case{a^{3}}{3\pi}{\mathbf E}_{\text{inc},\bot},&\quad
{\mathbf M}_{\text{eff}}= -2\case{a^{3}}{3\pi}{\mathbf B}_{\text{inc},\|}.\label{dipolesdebethe}
\end{eqnarray}
 The symbols $\bot$ and $\|$ refer to the perpendicular and tangential components of fields, respectively.  These dipole terms are very similar to the induced dipole terms which appear in the well established Rayleigh-Mie theory of diffraction by a small perfectly conducting sphere: ${\mathbf P}_{\text{induced}}= a^{3}{\mathbf E}_{\text{inc}}$, ${\mathbf M}_{\text{induced}}= -\case{a^{3}}{2}{\mathbf B}_{\text{inc}}$. The principal difference arises in the substitution
$2\rightarrow \case{1}{2}$ between the magnetic and electric dipole. Strong analogies exist between the theory of diffraction by small hole and the theory of diffraction by  small particles. We can understand the physical origin of these similitudes using the Clausius Mossoti formula. Indeed, following these formula, the polarization produced by a locally constant electromagnetic field ${\mathbf E}_{0}$, ${\mathbf B}_{0}$ in a small dielectric sphere of constant permittivities $\epsilon,\mu$ which is immersed in a homogenous medium with permittivities $\epsilon_{0},\mu_{0}$ is defined by: ${\mathbf P}=\epsilon_{0}\frac{\epsilon/\epsilon_{0}-1}{\epsilon/\epsilon_{0}+2}a^{3}{\mathbf E}_{0}$, and ${\mathbf M}=\frac{1}{\mu_{0}}\frac{\mu/\mu_{0}-1}{\mu/\mu_{0}+2}a^{3}{\mathbf B}_{0}$. With the condition  $\epsilon/\epsilon_{0}, \mu_{0}/\mu\ll 1$ for a hole in a perfect metal, we obtain the following two relations: ${\mathbf P}=-\frac{\epsilon_{0}}{2}a^{3}{\mathbf E}_{0}$, ${\mathbf M}=\frac{1}{\mu_{0}}a^{3}{\mathbf B}_{0}$ which are related by the same factor of 2 as for the Bethe case. In the opposite case, namely if  $\epsilon/\epsilon_{0}, \mu_{0}/\mu\gg 1$, we obtain the Rayleigh-Mie's solution. \\
The diffraction of light by a small aperture in a perfectly conducting plane can be qualitatively regarded as the complementary case to the scattering of a wave by a small conducting particle.
We can then consider that the incident light ``sees'' the hole as a (${\mathbf P}$,${\mathbf M}$) combination.
The general form of Eq.~\ref{dipolesdebethe} is certainly  conserved for various quasi circular opening in planar geometry and must essentially be independent on the aperture shape of subwavelength holes. As a good staring point we can anticipate the generality of this result, and we can admit that in any problem of diffraction by subwavelength holes in non planar screens
  the emission zone can be described by dipoles terms similar to Eq.~\ref{dipolesdebethe}. Nevertheless, numerous differences certainly arise from boundary effects when we look at the field near the metallic surface. In particular, the dipole orientations may change with the screen geometry as we will show later on.\\
%
%
\emph{Model}~-~In the case of an NSOM fiber tip, the aperture is situated on the apex of a conical, metal-coated structure (see Fig.~\ref{fig2}).
For eigenvalue-problems of conical symmetry around some origin, it is
convenient to express the fundamental solutions in spherical
coordinates $\left(r,\theta,\phi\right)$. The separation of the angular and radial variables in the Helmholtz
equation follows the ``Hankel(r)$\cdot$spherical-harmonics($\theta, \phi$)'' solution\cite{Jackson}
$\psi\left({\mathbf r}\right)=h_{\nu}\left(kr\right)\cdot Y_{\nu,m}\left(\theta,\phi\right)$ with the modification that in this
geometry the parameter $\nu$ can have a non-integer value which is determined by the boundary conditions.
 In analogy to the method of
multipole expansions with spherical harmonics used in free space, we can develop our propagating field as an expansion in electric
and magnetic ``quasi-multipoles''. The principal difference to the classical multipole expansion\cite{9} consists in the substitution $l\rightarrow \nu$.\\
We find that in general, a propagating electromagnetic field in conical geometry produces an angular distribution of radiated power of the form
\begin{eqnarray}
\lefteqn{\frac{d P}{d\Omega}= \frac{c}{8\pi k^{2}}\cdot \|\left(-i\right)^{\nu_{E}+1}
\sum _{\nu_{E},m}  a_{\nu_{E},m}^{\left(E\right)}
.\case{{\mathbf L}Y_{\nu_{E},m}\left( \theta,\phi\right)}
{\sqrt{\nu_{E}  \left( \nu_{E} +1 \right)}}\times{\mathbf e}_{r}  } \nonumber \\
 & & {} +\left(-i\right)^{\nu_{M}+1} \sum _{\nu_{M},m}  a_{\nu_{M},m}^{\left(M\right)}
\cdot\case{{\mathbf L}Y_{\nu_{M} ,m} \left( \theta,\phi\right)}
{\sqrt{\nu_{M} \left( \nu_{M} +1 \right)}}   \|^{2},
\end{eqnarray}
  with ${\mathbf L}=\frac{1}{i}{\mathbf r}\times {\mathbf \nabla}$.
Here, $\nu_{E}$ and $\nu_{M}$ refer to the electric (TM) and
 magnetic (TE) quasi-multipoles, respectively.
 The coefficients $a_{\nu ,m}$ depend on charge and current distribution
 on the metallic cone. It can be shown that for two roots $\nu$ and $\nu'$ (with $\nu \leq \nu' $), the ratio
 $a_{\nu'}/a_{\nu}$ is comparable to $\left(a/\lambda\right)^{\nu'-\nu}\ll 1$ which can be neglected. Therefore, we can omit all terms in the expansion except for the first two roots $\nu_{E,0}$, and $\nu_{M,0}$.\\
%
%
Specific boundary conditions can be written for a good metal such as aluminum.
They depend only on its
electric permittivity $\epsilon\left(\omega\right)$ ($ \simeq-34.5+8.5i$ for $\lambda=488$ nm). If we
neglect the skin depth $\delta\left(\omega\right)$ we arrive at the following boundary conditions on the cone:
 $\partial Y_{\nu_{M} ,m}\left( \theta,\phi\right)
/\partial \cos{\theta}\vert_{S}\simeq 0$, and $ Y_{\nu_{E} ,m}\left( \theta,\phi\right)\vert_{S}\simeq 0$.
However, we can make a more realistic approximation and use a finite value for the skin depth in the optical domain,
($\delta\simeq 10$ nm). In this case, there exist just outside the surface
tangential ${\mathbf E}_{\|}$ and  ${\mathbf B}_{\|}$ linked by the relation
${\mathbf E}_{\|}= \sqrt{\mu/\epsilon}{\mathbf B}_{\|}\times{\mathbf n}$. In this equation ${\mathbf n}$ is inwardly
 directed normal on the metal surface and $\sqrt{\mu/\epsilon}$ can be interpreted as a surface impedance. This relation determines an eigenvalue condition on $\nu$ and authorizes an infinite but discrete
 number of solutions which depend only of the tip angle
 $\beta$. For a value of $\beta$ on the order of $10^{\circ}$--$15^{\circ}$, we obtain
 $\nu_{E,0}\simeq 0.95$ and $\nu_{M,0}\simeq 0.96$. The $\nu$ values
and, consequently, the angular profile are not sensitive to variations of the tip angle $\beta$ unless it exceeds
$20^{\circ}$.
\\
%
%
It is important to note that the boundary conditions cannot be applied inside a sphere of radius $a/\left(\sin{\beta}\right)$ because it contains no metal. This so-called aperture zone
 represents the ``terra incognita'' near-field zone ($r/\lambda\lesssim 1$) of the fiber tip (cf. Fig.~\ref{fig2}).
%
%
The coefficients $a_{\nu ,m}$ can be approximated by a Taylor series around the tip origin. $m=\pm 1$ are the only authorized terms for linearly polarized light propagating down the fiber. These terms possess the form $a^{E}\sim {\mathbf P}_{eff}\cdot{\mathbf E}_{0}$, $a^{M}\sim {\mathbf M}_{eff}\cdot{\mathbf B}_{0}$ which depend on the field ${\mathbf E}_{0},{\mathbf B}_{0}=\hat{{\mathbf z}}\times{\mathbf E}_{0}$ at the aperture and on the two dipoles:
  \begin{eqnarray}
{\mathbf P}_{\textrm{eff}}=\frac{1}{4\sqrt{2}}\frac{a^{3}}{\sin{\beta}^{3}}{\mathbf E}_{0} ,\quad&
{\mathbf M}_{\textrm{eff}}=2\hat{{\mathbf z}}\times{\mathbf P}_{\textrm{eff}}.
\end{eqnarray}

The normalized angular distribution of the radiated power can then be expressed as
  \begin{equation}
\displaystyle
\frac{1}{P}\frac{d P}{d\Omega}\simeq\frac{2}{5}\left\|
\textrm{Im}\frac{{\mathbf L}Y_
{\nu_{E} ,1}\left( \theta,\phi\right)}{\sqrt{\nu_{E} \left( \nu_{E} +1 \right)}}
  \times {\mathbf e}_{r} +2\textrm{Re}\frac{{\mathbf L}Y_{\nu_{M} ,1}\left( \theta,\phi\right)}
{\sqrt{\nu_{M} \left( \nu_{M} +1 \right)}} \right\|^{2}.
\label{epsilon}
\end{equation}
In this formula Re and Im are related to the real and imaginary parts of the generalized harmonic functions. It appears also a
characteristic factor $2$ between the electric and the magnetic quasi-multipole. It is important to remark that
neither the absolute value of the radiated power nor the absolute value of the dipoles are described by our model.
This would involve a much more elaborate description of the core/cladding characteristics inside the fiber tip.
We can also note that the normalized profile does not depend on the aperture radius.
  Long wavelength light is blind to subwavelength details, and therefore the aperture size
 does not constitute a critical parameter of our model.\\
%
%
Eq.~\ref{epsilon} reduces to the
Oberm\"uller-Karrai model when we put $\nu=1$ (a value close to the real values $\nu_{E,0}$ and $\nu_{M,0}$). This corresponds to the ideal case of a
linear, radiating two-dipole antenna. In this limit $\beta$ and $a/\beta$ tend to zero, and we can write
\begin{equation}
\frac{1}{P}\frac{d P}{d\Omega}^{\nu=1}= \frac{3}{8\pi}\cdot\frac{\| \left( {\mathbf e}_{r}\times
{\mathbf P}_{\text{eff}}\right)\times{\mathbf e}_{r} +{\mathbf M}_{\text{eff}} \times {\mathbf e}_{r} \|^{2}}{ \|{\mathbf P}_{\text{eff}}\|^{2} +
\|{\mathbf M}_{\text{eff}}\|^{2} }.
\label{theta}
\end{equation}
The effective electric and magnetic dipoles are linked by
${\mathbf M}_{\text{eff}}=2{\mathbf e}_{z}\times{\mathbf P}_{\text{eff}}$ and by the generalized Bethe relation
${\mathbf P}_{\text{eff}}\propto a^{3}{\mathbf E}_{\text{inc}}$ where ${\mathbf E}_{\text{inc}}$ denotes the incident light polarization. Also, the factor of 2 reappears which is common to all cases of diffraction by a subwavelength aperture as precedently explained.
The two-dipole model has another significance because the general expression Eq. \ref{epsilon} converges to Eq. \ref{theta} for small azimuthal angles $\theta$. As a direct consequence the phenomenological Oberm\"uller-Karrai's model is justified and constitute a good approximation valid for small azimuthal angles. \\
%
%
\emph{Results and discussion}~-~The following Figs.~\ref{fig3},\ref{fig4} show experimental data for both S- and P-polarization and compare them with the calculated profiles of our model. Also shown are the curves for the simpler two-dipole model which is a good approximation only for smaller angles. There are no adjustable parameters in both models.

As shown in Fig.~\ref{fig3} (S-polarization), our model and the Oberm\"uller-Karrai model are equivalent for the experimentally accessible angles. They follow the experimental data for all recorded angles. This behavior is different for
the P-polarization as shown in Fig.~\ref{fig4}. Our model predicts an important ``backscattering'' in good agreement with the experiment. The two-dipole model does not correctly predict this effect. Qualitatively, the difference between both polarisations can be understood by realising that the Poynting lines in the P-polarisation are tangential to the metal surface, and are consequently very sensitive to boundary conditions on this surface, while in the other polarisation, Poynting lines tend to avoid the metal surface because both the electric and magnetic fields vanish there. The quasi-multipole model is nevertheless in disagreement
 with the measured data for extreme $\theta$ values. We believe that this discrepancy is due to several effects. First, there are strong near-field effects on the surface of the metallic cone which are not accounted for in the model. They appear in higher order terms in the quasi-multipole expansion (these terms are not considered here). Additionally,  incoherent scattering due to surface roughness reduces the collected light for large angles which could explain the discrepancy.\\

%
%
\emph{Conclusion}~-~In summary, we have developed an analytical model for the far-field transmission pattern of conical, metal-coated fiber tips with a small, subwavelength aperture. It contains, as a first approximation, the two-dipole model anticipated earlier\cite{6}. It justifies the generality of the Bethe's result for the case of a conical screen. Our model is in good agreement with the experimental data and
explains in particular the backward emission found for the P-polarization. The angular transmission profile does not depend on the aperture radius
for small holes, and is insensitive to the tip angle as long as it is smaller
 than $20^{\circ}$.

\begin{figure}
\includegraphics[width=10cm]{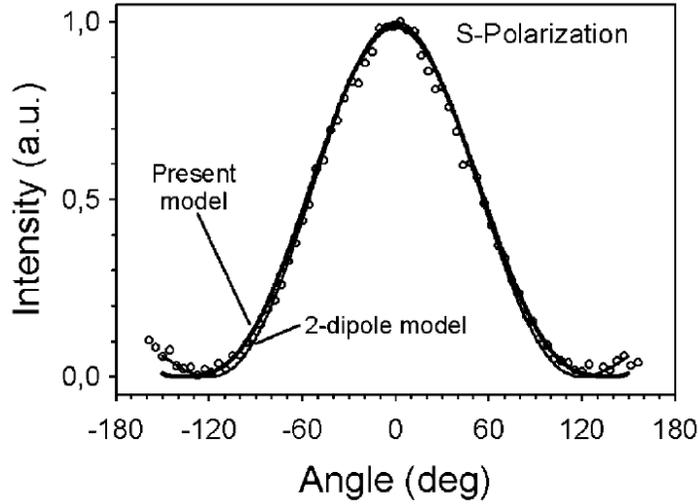}  \caption{Normalized angular distribution of radiated power in S-polarization where the detector is scanned
perpendicular to the plane of polarization of the incident light. Experimental data are shown in circles. The thick line corresponds to the present quasi-multipole model, the thin line to the 2-dipole model. The experimental data (aperture radius $a=30$ nm, light  wavelength $\lambda=633$ nm) are taken from Refs.~2,3.} \label{fig3}
\end{figure}

\begin{figure}
\includegraphics[width=10cm]{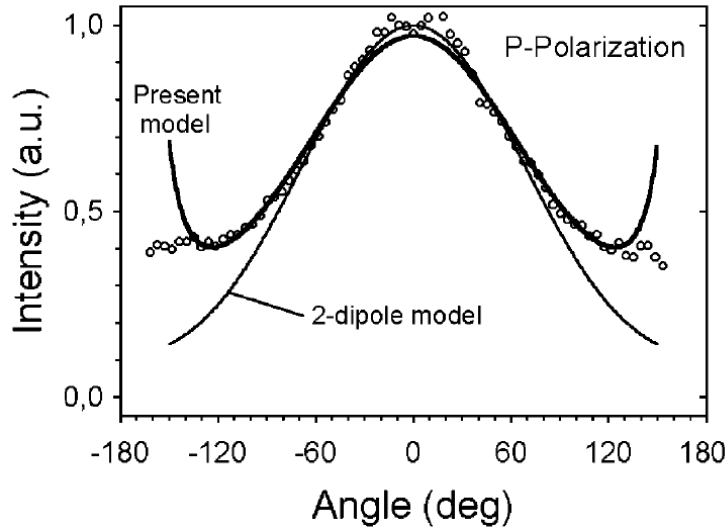} \caption{Same as Fig.~\ref{fig3} but for the P-polarization where the incident electric field is parallel to the analysis plane.} \label{fig4}
\end{figure}

\end{document}